\documentclass[aps,12pt,prd,superscriptaddress,amsfonts,amssymb,amsmath,eqsecnum
,nofootinbib,floatfix]{revtex4}
\usepackage[dvips]{graphicx}
\usepackage{psfrag}
\usepackage{latexsym,amssymb,amsmath}
\usepackage{euler}

\allowdisplaybreaks
\begin{document}
\title{Notes about Noise in Gravitational Wave Antennas Created by Cosmic
        Rays}
\author{V.B. Braginsky}
\affiliation{Physics Dept, Moscow State University, Moscow, 119992
        Russia}
\author{O.G. Ryazhskaya}
\affiliation{Inst. for Nuclear Research, prospect of 60-th Anniversary of
        October, 7-a, Moscow, 117312 Russia  }
\author{S.P.         Vyatchanin}
\affiliation{Physics Dept, Moscow State University, Moscow, 119992
        Russia}
\date{\today}

\begin{abstract}
We discuss three mechanical effects initiated by cosmic rays
which may limit the sensitivity of gravitational wave antennas.  Unsolved
problems are formulated and several recommendations for the antenna designs
are presented.

\end{abstract}

\maketitle

\section{Introduction}

Since 1969 several groups of researchers have presented results of the
analysis and estimates of possible contribution of cosmic rays to the noise
floor of gravitational wave antennas (both bar antennas and antennas based on
free masses, e.g. \cite{BeHo69, ChMi91} ). The currently achieved
sensitivities in Initial LIGO project (see \cite{BaWe99, Br00}) and the
planned sensitivity in the next stage (Advanced LIGO which is planned to
operate within a few years from now) are respectively $h\simeq 10^{-21}$ and
$h\simeq 10^{-22}$ for the amplitude of the perturbation of the metric (at the
mean frequency $f\simeq 100$~Hz in the bandwidth $\simeq 100$~Hz).
Independently of Advanced LIGO several groups of researches which belong to
the LSC (LIGO Scientific Collaboration) are continuing the analysis of new
topologies and designs of the antennas which will permit them to reach a
sensitivity even better than in Advanced LIGO (see e.g. \cite{Br00, Kh02}).

Apart from this activity during the last two decades there is substantial
progress in the collection of data and in the resolution in measuring features
of cosmic ray showers (cascades) (see e.g. \cite{En88,Ry96,Mu03}).
Thus it is reasonable to revise the contribution of cosmic rays to the noise in
gravitational wave antennas.

In the present note we limit ourselves to only three possible
mechanical ``actions'' on the rest masses (mirrors):

\begin{enumerate}
\item Direct transfer of mechanical momentum from the cascade to the LIGO
mirror.

\item Distortion of the mirror's surface due to  the heating by the cascade
and subsequent thermal expansion --- thermoelastic effect.

\item Fluctuating component of the Coulomb force between electrically charged
mirror and grounded metal elements located near the mirror's surface.

\end{enumerate}

\section{Direct transfer of cascade mechanical momentum to LIGO antenna
        mirror}

In Advanced LIGO it is planned to use nearly cylindrical mirrors in the main
Fabry-Perot (FP) optical resonator (mirror's diameter -- $2R\simeq 35$~cm,
height -- $H\simeq 20$~cm, total mass of fused silica ($SiO_2$) mirror
$M=40$~kg). The mirrors will be suspended on $SiO_2$ fibers in a horizontal
vacuum tube. The axis of the mirrors will be parallel to the tube. The
distance between mirrors will be $L\simeq 4$~km. Two mirrors have to respond
to the gradient of acceleration generated by a gravitational wave propagating
in a perpendicular direction to the FP axis.

\begin{table}[h]
\caption{The parameters of high energy cascades we used for estimates.  ${\cal
E}$ is cascade energy, $J_\mu, J_h, J_e$ are the fluxes of cascades produced
by muons, hadrons and by soft component, consequently, at sea level;
$N_{e,\, max}$ is a number of electrons in the cascade maximum;
$\Delta{\cal E}$ is the energy lost by the cascade in the 20 cm of $SiO_2$;
$N_{ev}$ is the expected number per year of events with energy losses higher
than $\Delta {\cal E}$. }

\label{tab1}
\begin{tabular}
{|p{0.15\textwidth} | p{0.14\textwidth} |p{0.14\textwidth}  |p{0.14\textwidth}
 |p{0.15\textwidth}  |}
\hline
${\cal E}$, TeV & 0.5 & 1 & 2 & Ref\\
\hline
$J_\mu\, 1/\text{cm}^2\text{s}$& $1.8\times10^{-9}$ & $2.8\times 10^{-10}$ &
 $4.3\times 10^{-11}$&\cite{En88,Ry96,Ry05,Vo79}\\
$J_h\, 1/\text{cm}^2\text{s}$& $2.5\times10^{-9}$ & $4.0\times 10^{-10}$ &
 $7.2\times 10^{-11}$&\cite{Mu03}\\
$J_e\, 1/\text{cm}^2\text{s}$& $3\times10^{-10}$ & $8\times 10^{-11}$ &
 $1.7\times 10^{-11}$&\cite{Ha69}\\
$N_{e,\, max}$ & 1000 & 2000 & 4000& \\
$\Delta {\cal E}$, GeV &60 &120 & 230& \\
\hline
$N_{ev}$ 1/year& $\sim 110$&20&$3\div4$& \\
\hline
\end{tabular}
\end{table}

It is reasonable to expect that a certain part of all cascades (showers)
generated by very high energy muons of cosmic rays will pass through
the mirror along axes which have a small angle with the FP resonator
axis.

Thus one may expect that not a very small
fraction of a cascade with energy $\cal E$ will be lost in the ``travel''
through $20$~cm of fused silica mirror. This fraction $\Delta{\cal E}$
will give a momentum $\Delta P=\Delta {\cal E}/c$ to the mirror along
its axis ($c$ is speed of light). Correspondingly this mirror position
will  change during time $\tau$ by a value
$$ \Delta x\simeq \frac{\Delta
{\cal E}\tau}{Mc} , \quad \tau\simeq 0.01 \ \text{s}
$$

The narrow cascades in the direction close to  vertical
can be produced by so-called unaccompanied hadrons \cite{Mu03} and
electron-photon (soft) component of cosmic rays \cite{Ha69,Be48}. High
energy $({\cal E}>0.5 TeV)$ cosmic ray muons are the origin of electromagnetic
and nuclear cascades. These particles are born in decay of $\pi^\pm$-
and $K^\pm$ -mesons and charmed particles generated in  hadronic
showers. The showers are initiated by interactions of primaries with
the nuclei of air at high altitudes in the atmosphere. The energy
spectra of all particles (and all cascades generated by them) have
power law shape $F({\cal E}_c) \sim {\cal E}_c^{-\gamma}$, $\gamma$ is
the power index, $\gamma=2.73\pm0.05$ for $J_\mu$ \cite{Ry96};
 $\gamma=2.5\pm0.1$
for $J_h$ \cite{Mu03}; and $\gamma=2.2\pm0.2$ for $J_e$ \cite{Ha69}. For the
 estimation of the
parameters of cascades it is possible to use a well elaborated model of
this type of cascades \cite{Be48}.

In Table~\ref{tab1} we present numerical estimates for ``an appropriate
candidate''  for the discussed process. For several energies $\cal E$ of
cascade one can find from the literature the fluxes of cascades produced by
muons ($J_{\mu\text{ und}}$) underground \cite{En88,Ry96}, hadrons ($J_h$)
 \cite{Mu03}
and by the soft component ($ J_e$) \cite{Ha69} at  sea level  and calculate
the mean number $N_{e,\, max}$ of electrons in the cascade maximum. The flux
of cascades produced by muons at sea level ($J_\mu$) was estimated from
experimental data obtained underground ($J_{\mu\text{ und}}$) \cite{En88}
taking into account the calculations of energy spectrum and angular
distribution of muons at sea level \cite{Ry05, Vo79}. The energy lost
$\Delta\cal E$ of a cascade was obtained by integration of energy losses of
electrons in the mirror taking into account the energy spectrum of electrons.
The expected number $N_{ev}$ of events per year with energy losses equal to or
higher than $\Delta {\cal E}$  was calculated by the formula
$N_{ev}=[J_\mu\times S_1+(J_h+J_e)\times S_2]\times T$, ($S_1, S_2$ are the
areas of the mirror perpendicular and parallel to the axis, respectively,  $T$
is one year $T\approx 3\times 10^7$ s). These estimates were made under the
assumption that the cascade is coming into the mirror being well developed (more
than 6 interactions of the fastest electrons).

% A cascade with energy
%higher than 2 TeV created by muons, hadrons or soft component having power
%law shape energy spectrum will be passing through the mirror approximately $2
%\div 3$ times per year.

The values of $\Delta {\cal E}$, presented in Table \ref{tab1} permit us to
estimate the mirror's displacement $\Delta x\simeq (0.8\div 3)\times
10^{-18}$~cm. This numerical value of $\Delta x$ is less than the amplitude of
the sensitivity planned in Advanced LIGO: $\Delta L\simeq hL/2\simeq 2\times
10^{-17}$~cm.

Evidently one may expect a much more frequent rate of events: e.g. a
cascade with initial energy ${\cal E}=1$~GeV will fly through the same
mirrors several times per second. But because the value of $\Delta
{\cal E}$ in this case will be approximately $3$ orders smaller the net
effect will be negligible compared to the planned sensitivity of
Advanced LIGO.

\section{Distortion of the mirror's surface due to the thermoelastic effect and
heating by the cascade}

The lost in the mirror's bulk energy of the cascade is likely to be
distributed into two parts. The first one is the rise of free energy of the
solid (creation of new dislocations, of new clusters, etc) and the second part
produces direct heating of a narrow channel. There is no rigorous analysis in
the quantum theory of solids which permits us to get a reliable value of the
ratio of these two parts. But it is very likely that the second one (the
heating) will dominate.

The energy $\Delta\cal E$ is converted into heat over a length $H=20$~cm
(thickness of mirror). This heat will be produced in the trace of the cascade in
the volume $\pi R_c^2H$ where $R_c\simeq 1\div 7$~cm is the radius of the
cascade trace. Assuming that the volume $\simeq R_c^3$ on the mirror's surface
can freely expand we obtain an estimate of the height $\Delta H$ of the ``
hill'' with footprint $\simeq R_c^2$ on the surface due to thermal expansion:

\begin{align}
\Delta H \simeq \frac{R_c}{H}\times
        \frac{\Delta {\cal E}}{\rho C R_c^3}\times R_c\alpha
\end{align}
Here $\rho\simeq 2.3$~g/cm$^3$ is the density,
$C\simeq 7\times 10^6$~erg/g\,cm$^3$K is the heat capacity and
$\alpha\simeq 5.5\times 10^{-7}$~$K^{-1}$
is the thermal expansion coefficient of fused silica. The height $\Delta
H_\text{av}$ averaged over  laser beam spot with radius $r\simeq 10$~cm is
approximately equal to
\begin{equation}
\Delta H_\text{av}\simeq \Delta H \times \frac{R_c^2}{r^2}\simeq
        \left\{
        \begin{array}{cc}
        2\times 10^{-18}\, \text{cm}&\quad \text{if}\
                \Delta{\cal E}=60\, \text{GeV,}\  \text{if}\ R_c=1 \ \text{cm}
 \\
        5.4\times 10^{-17}\, \text{cm}&\quad \text{if}\
                \Delta{\cal  E}=230\, \text{GeV,}\ \text{if}\ R_c=7 \ \text{cm}
        \end{array}
        \right.
\end{equation}

The  displacement $\Delta H_\text{av}$ of the surface considered above is
produced by a cascade developing mainly perpendicular to the surface of the
mirror. However, there are cascades having traces approximately parallel to
the surface. Such a ``parallel'' event produces a greater contribution to
the fluctuational displacement of the mirror's surface:
\begin{align}
\Delta H_\text{av,parall}&\simeq
        \frac{\Delta {\cal E}}{\rho C R_c^2 2r}\times R_c\alpha
        \times \frac{R_c}{r}\simeq
        \left\{ \begin{array}{cc}
                2\times 10^{-17}\,\text{cm}&\quad
                \text{if}\ \Delta{\cal E}=60\, \text{GeV} \\
                7.7\times 10^{-17}\, \text{cm}&\quad
                \text{if}\ \Delta{\cal  E}=230\, \text{GeV}
        \end{array} \right.
\end{align}

However, such events are rare than
``perpendicular'' ones by a factor of  about $ R_c^2/r^2\simeq 0.01\div 0.5$
if we roughly assume the  spherical symmetry of showers
distribution\footnote{For
more accurate consideration we have to take into account the real spartial
distribution of muon \cite{Ry05}, hadron \cite{Mu03} and soft \cite{Ha69}
components of cascades.}.
%for $0.5$ TeV cascade it  takes place approximately 1 times per year and for
%$2.5$ Tev cascade --- $2 \div 3 $ times per 100  year.

\section{Fluctuating Coulomb force}

In 1995 R.~Weiss pinpointed the potential danger from electrical charge
accumulated on the mirror's surface \cite{We95}. Direct measurements of
the values of electrical charge density $\sigma$ on models of mirrors were
performed independently by several groups \cite{Mo03,Mi02,Mi04}.  In these
measurements it was demonstrated that the values of $\sigma$ on
models of mirrors (fabricated from fused silica) was from $10^6$ to $10^7$
electrons per cm$^2$ and in several cases even higher.

Recently  V.P. Mitrofanov and his colleagues \cite{Mi02,Mi04} have measured
the values of $\sigma$ in the same vacuum chamber in which record-high
quality factors of pendulum and violin modes ($Q>10^8$) were demonstrated.
These measurements were performed during several months and a slow, long
lasting drift of $\sigma$ was observed. The high values of surface density of
charge obtained mean that the electrostatic potential of the mirror may
exceed $100$~V.

V.P. Mitrofanov and his colleagues \cite{Mi02} have discovered a monotonic
rise of negative charging --- $d\sigma /dt\simeq 10^5$ electrons per cm$^2$
per month. This effect can be qualitatively explained by the model of
transition effect occurring during transition of cascade particles, soft
component and gamma-quanta of natural radioactivity through the iron
``envelope'' to the fused silica mirror (i.e. vacuum chamber).

In the cosmic rays of low energy there are more electrons than positrons.
 Compton effect, photoelectric effect and the process of ionization are the
 sources of the excess. In electromagnetic cascades the number of particles with
 low energy is much larger than the number of particles with high energy. The
 number of particles having energy less than 0.05 of the critical energy in the
 matter (${\cal E}_{cr,Fe}=20,7 MeV$ for iron, ${\cal E}_{cr,Al}=40 MeV$ for
 aluminium, ${\cal E}_{cr,SiO_2}=47,3 MeV$ for $SiO_2$) is equal to $\simeq
 20\%$, and the number of particles having energy less than 0.02 of the critical
 energy in the material is equal to $\simeq 10\%$ of number of particles in the
 cascade maximum. The cosmic rays are very sensitive even to  thin layers of
 matter. If a cascade developed in the heavy material comes to the
material consisting of lighter atoms, it brings to the light material more
electons than it takes away. This is explained by the fact that the number
$N_e$ of electrons produced in the material is proportional to the fraction
$N_e\sim {\cal E}/{\cal E}_{cr}$, where ${\cal E}$ is the cascade energy.

As an example we consider a cascade in the maximum of its development coming
from  iron to fused silica.
%$=1 $~MeV
In this case cascade theory gives the formulas for the number of low energy
 electrons produced in iron and in fused silica:
 \begin{align}
N_{e,Fe}( E < 1 \text{MeV}) &= 0.2\times
        \frac{0.2}{\sqrt {\ln({\cal E}/E_{cr,Fe})}}\times
        \frac{{\cal E}}{{\cal E}_{cr,Fe}},\quad
	(1\, \text{MeV}=0.05\,{\cal E}_{cr,Fe}),\\
N_{e,SiO_2}( E < 1 \text{MeV})& = 0.1\times
        \frac{0.3}{\sqrt {\ln({\cal E}/{\cal E}_{cr,SiO_2})}}\times
        \frac{{\cal E}}{{\cal E}_{cr,SiO_2}},
	\quad 	(1\,\text{MeV} =0.021\,{\cal E}_{cr,SiO_2}).
\end{align}
So we see that the number of low energy electrons produced in iron and coming to
the  mirror is about $3$ times larger than the number produced in fused silica
and  outgoing from it:
$$
\frac{N_{e,Fe}}{N_{e,SiO_2}}\approx\frac{0.4\,{\cal E}_{cr,SiO_2}}{0.3\,{\cal
 E}_{cr,Fe}}
\approx 3
$$

This electron excess will stay near the surface of the mirror and will give an
additional charge to it.  The estimates of the number of electrons with energy
less than 1 MeV coming from iron to fused silica are given in the
Table~\ref{tab2}. This effect can qualitatively explain the monotonic rise  of
{\em negative} charge observed in \cite{Mi02,Mi04}. However, for quantitive
explanation a detailed analysis has to be performed.

\begin{table}[h]
\caption{The mean number N of electrons with energy less than 1 MeV coming
	from iron to fused silica.}\label{tab2}
\begin{tabular}
{|p{0.15\textwidth} | p{0.14\textwidth} |p{0.14\textwidth}  |p{0.14\textwidth}
 |}
\hline
${\cal E}$, TeV & 0.5 & 1 & 2 \\
\hline
$N(E < 1 \text{MeV}) $ & 450 & 900 & 1700 \\
\hline
\end{tabular}
\end{table}

The initial design of the ``entourage'' of the suspended mirror
includes several parts which are planned to be made of metal. These
parts include a ``cradle'' situated under mirror (this `` cradle'' has
to catch the mirror if one or all fibers break). Other parts are called
``stoppers'' which have to limit large horizontal swings of the mirror
in case of an earthquake. These parts have to be grounded. Thus due to
electrical charging of the mirror it is very reasonable to expect that a
d.c. Coulomb force may act on the mirror. If a grounded metal part has
a flat surface $S$ that is close to a part of the mirror's surface, then $$
F_{dc}\simeq 2\pi S \sigma^2\simeq  1.5\times 10^{-2}\,
\text{dyn},\quad \text{if}\  S=10^2\, \text{cm}^2 $$

This numerical estimate of the d.c. force has to be taken into account in
the design of feedback actuators which have to maintain the distance
between antenna mirrors with accuracy better that $\lambda/{\cal
F}\simeq 10^{-9}$~cm ($\cal F$ is the finesse, $\lambda$ is the optical wave
length). More important is another effect: The a.c. component of the
Coulomb force may mimic the force $F_\text{grav}$ that antenna has to
register:
\begin{align} F_\text{grav}\simeq
\frac{hLM\omega_\text{grav}^2}{2}\simeq 3\times 10^{-7}\ \text{dyn}
\end{align}
A comparison of the  values  presented above indicates that a
relative fluctuation $\Delta \sigma/\sigma\simeq 10^{-5}$  of surface
charge density will inevitably produce a `` step'' of $F_\text{ac}$
approximately equal to the amplitude $F_\text{grav}$ which is the goal
of Advanced LIGO. Note that $N_{E<1\, \text{MeV}}$ electrons outgoing from iron
 to
fused silica presented in Table \ref{tab2} comes to the square about
$\pi R_c^2\simeq 100\ \text{cm}^2$ ($R_c$ is the  radius of cascade) .
Then one can estimate the  relative fluctuations of charge density
$\Delta \sigma /\sigma$ caused by a single cascade with energy ${\cal
E}= 2$~TeV:
$$
\frac{\Delta \sigma}{\sigma}\simeq 10^{-6}\div 2\times 10^{-5}
$$
We see that this fluctuation is strong enough to produce an a.c. component of
Coulomb force larger than $F_\text{grav}$ if a relatively large surface of
grounded metal plate will be located near the mirror.
%This change corresponds to $\sigma \simeq 10^2$
%electron per cm$^2$ and the total number of ``necessary'' electrons
%$S\Delta\sigma\simeq 10^4$.

\section{Conclusion}

It is evident that the first two effects, being not very strong ones for
Advanced LIGO, may be relatively  easily vetoed by requiring coincidence
between detectors (if at least two antennas are operating). On the other hand,
the veto  can not be considered as an absolute ``remedy'' for low values of
the signal to noise ratio. It is worth noting that in the next stage after
Advanced LIGO these two effects will make serious contributions to the
level of the noise floor.

This conclusion may not be automatically extended to the third effect. First
of all because the negative charging may be high when the mirrors ``spend'' a
long time in the vacuum (an year or longer) without scheduled removal of the
accumulated electrical charge. The second ``reason'' is the design of all
metal parts of the mirror's ``entourage''. One recommendation for this design
is evident: it is necessary to use small square areas of all metal element
that are near the mirror's surface and to place these elements as far away as
possible.

There are two evident recommendations for the consequent measurement and
analysis:
\begin{enumerate}
\item To measure bursts of electrons which appears on the mirror's surface
with resolution better than $10^2$ e/cm$^2$ and time shorter than $10^{-2}$
sec.
\item To analyze the possibility to cover the mirror's surface over the
coating with a transparent few nanometers thick  layer with substantial
conductivity to reduce the d.c. component of electrical charge.
\end{enumerate}

%The estimates presented above are based on simple model of one type of the
%cascade. Other types (including those with hadrons) also deserve at least a
%similar analysis.

In all three effects considered above the mechanical action on the
mirror produces a step-like displacement (either of the mirror's center of
mass or of its surface). This type of response is similar to the one
predicted for the shape of gravitational wave bursts created in the process
of supernova explosion predicted by V.Imshennik \cite {Im98}.  The
predicted rate of these bursts is approximately two orders higher than
the rate of neutron star merger events.

%\subsection*{Acknowledgements}
\acknowledgments

We are very  grateful to V.~Mitrofanov, Ph.~Willems and T.~Roganova for fruitful
 stimulating discussions. This work was supported by the LIGO team from
Caltech and in part by NSF and Caltech grants PHY0098715 and PHY-0353775, by
the Russian Agency of Industry and Science, contracts No.\, 40.02.1.1.1.1137
and No. \,40.700.12.0086, by the Russian Foundation of Fundamental
Research, grants 03-02-16975-a, 03-02-16414, SSchool-1782.2003.2 and Program for
 Fundamental Research of Presidium of RAS ``Neutrino Physics'' 2005.


\begin{thebibliography}{10}

\bibitem{BeHo69} B.L. Beron and R. Hofstadter, Rhys. Rev. Lett. {\bf 23} (4),
184 (1969).

\bibitem{ChMi91} J. Chiang and P. Michelson, Nuclear Instruments and Methods in
Physics Research {\bf A311}, 603 (1991).

\bibitem{BaWe99} B. Barish and R. Weiss, Physics Today, {\bf 52} (10), 44
(1999)

\bibitem{Br00} V.B. Braginsky, Physics ``Uspechi'', {\bf 43} (7), 691 (2000).

\bibitem{Brb00} V.B. Braginsky, M.L. Gorodetsky, F.Ya. Khalili and K.S. Thorne ,
 Phys. Rev. {\bf D61}, 4002 (2000).

\bibitem{Kh02} F.Ya. Khalili, Phys. Lett. {\bf A298}, 308 (2002).

\bibitem{En88} R.I. Enikeev, G.T. Zatsepin, E.V. Korolkova, V.A.
Kudryavtsev,  A.S.~Malguin, O.G.~Ryazhskaya, F.F.~Khalchukov, Sov. J. Nucl.
Phys.,         {\bf 47} (4), 1044         (1988).

\bibitem{Ry96} O.G. Ryazhskaya,  Il Nuovo Cimento, {\bf 19c} (5), 655, (1996).

\bibitem{Mu03} M. M\"uller, T. Antoni, W.D. Apel {\em et al} for KASCADE
 collaboration, proc. of 28th ICRC, Tsukuba, vol. 1, 101, (2003)

\bibitem{Ha69} S. Hayakawa, Cosmic Ray Physics, p I, Edited by R.E.~Marskak, New
 York, J. Wiley \& Sons, (1969)

\bibitem{Be48} S.Z. Belenkiy, Cascade Processes in Cosmic Rays, M.-L., (1948),
 in Russian

\bibitem{Ry05} O.G. Ryazhskaya, L.V. Volkova,
        Nuclear Physics {\bf B143}, 527 (2005).

\bibitem{Vo79}   L.V. Volkova, G.T. Zatsepin, L.A. Kuzmichev, Sov. J. Nucl.
 Phys., {\bf 29}, 645 (1979), Preprint FIAN, N72 (1969)

\bibitem{We95} R. Weiss, %Note on Electrostatic in the LIGO Suspension,
        LIGO document T960137-00E.
        Available in  {\sf http://admdbsrv.ligo.caltech.edu/dcc/}.

\bibitem{Mo03} M.J. Mortonson, C.C. Vassiliov, D.J. Ottaway, D.H.~Shoemaker and
G.M.~Harry, Rev. Sci. Inst. {\bf 74} (11) 4840
(2003).

\bibitem{Mi02} V.P. Mitrofanov, L.G. Prokhorov and K.V. Tokmakov, Phys. Lett.
{\bf  A300}, 370
(2002).

\bibitem{Mi04} V.P. Mitrofanov, L.G. Prokhorov,  K.V. Tokmakovand, Ph. Willems,
 Class. Quantum Grav. {\bf 21} (5),
1083 (2004).

\bibitem{Im98} V.S. Imshennik, I. Popov, Astron. Lett. {\bf 24}, 206 (1998).


\end{thebibliography}
\end{document}